\documentclass[12pt, letterpaper, notitlepage]{iopart}

\usepackage[usenames]{color}
\usepackage{graphicx}
\usepackage{cite}
\usepackage{subfigure}
\usepackage{hyperref}

\newcommand{\bk}{{{\bf{k}}}}

\newcommand{\bQ}{{{\bf{Q}}}}

\newcommand{\bvr}{{{\bf{r}}}}

\newcommand{\ua}{\uparrow}
\newcommand{\da}{\downarrow}
\newcommand{\bS}{{\vec{S}}}

\newcommand{\HH}{\mathcal{H}}

\begin{document}
\title{Nematic and spin-charge orders driven by hole-doping a charge-transfer insulator}
\author{Mark H. Fischer$^{1,2}$, Si Wu$^3$, Michael Lawler$^1$, Arun Paramekanti$^{3,4}$, Eun-Ah Kim$^1$}
\address{
$^1$ Department of Physics, Cornell University, Ithaca, New York 14853, USA}
\address{
$^2$ The Weizmann Institute of Science, Rehovot 7610001, Israel}
\address{
$^3$ Department of Physics, University of Toronto, Toronto, Ontario M5S 1A7, Canada}
\address{
$^4$ Canadian Institute for Advanced Research, Toronto, Ontario M5G 1Z8, Canada}

\begin{abstract}
Recent experimental discoveries have brought a diverse set of broken symmetry states to the center stage of research on cuprate superconductors. Here, we focus on a thematic understanding of the diverse phenomenology by exploring a strong-coupling mechanism of symmetry breaking driven by frustration of antiferromagnetic order. We achieve this through a variational study of a three-band model of the CuO$_2$ plane with Kondo-type exchange couplings between doped oxygen holes and classical copper spins. Two main findings from this strong-coupling multi-band perspective are 1) that the symmetry hierarchy of spin stripe, charge stripe, intra-unit-cell nematic order and isotropic phases are all accessible microscopically within the model, 2) many symmetry-breaking patterns compete with energy differences within a few meV per Cu atom to produce a rich phase diagram. These results indicate that the diverse phenomenology of broken-symmetry states in hole-doped antiferromagnetic charge-transfer insulators may indeed arise from hole-doped frustration of antiferromagnetism.
\end{abstract}

\section{Introduction}
Translational-symmetry breaking in hole-doped La-based nickelates and cuprates in the form of  static spin and charge stripes
has been well established for almost two decades\cite{PhysRevLett.73.1003,tranquada:1995,tranquada:1996,abbamonte:2005vn}. However, until recently,
the ubiquity of such phenomena had not been clear. The recent surge of experimental discoveries reporting spin or charge order in all families of hole-doped cuprates\cite{Comin24012014,daSilvaNeto2014,Shekhter:2013uq,blackburn:2013,chang:2012,achkar:2012,PhysRevB.86.020504,ghiringhelli:2012,Wu:2011kx,okamoto:2010,lawler:2010,Daou:2010,gull:2009,Li:2008fk,PhysRevB.78.020506,PhysRevLett.100.127002,hinkov:2008} and even in some Fe-based superconductors \cite{Kasahara:2012ys} has brought the diverse phenomenology of broken-symmetry states to the forefront of study of high T$_c$ superconductors. In particular there is extensive indication that ${\bf Q}=0$ intra-unit-cell (IUC) orders \cite{lawler:2010,Daou:2010,PhysRevB.86.020504,Li:2008fk,PhysRevB.78.020506} and (short-ranged) ${\bf Q}\neq0$ modulations \cite{Comin24012014,daSilvaNeto2014,blackburn:2013,chang:2012,achkar:2012,ghiringhelli:2012} coexist. 
In this new landscape of ubiquitous and diverse forms of broken-symmetry reports, an emerging central question is whether the diverse set of phenomena share a common origin 
or if each phenomenon should be studied on its own.

Closely linked to the question of whether a thematic understanding of the observed phenomena is attainable is a theoretical question of whether to take the weak-coupling Fermi-surface instability  perspective or to take the strong-coupling perspective. From a Fermi-surface instability perspective, the simultaneous occurrence of multiple orders requires fine tuning as one usually finds a single dominant instability in one ordering channel. For instance, it was shown that antiferromagnetic exchange interactions\cite{PhysRevLett.111.027202} or short-range repulsive interactions \cite{PhysRevB.88.155132} can drive an instability towards density modulation along an incommensurate vector ${\bf Q}=(Q_0,Q_0)$ with a dominantly d-form factor. Although a ${\bf Q}=0$ order with the d-form factor will be equivalent to IUC nematic order observed in various experiments\cite{lawler:2010,Daou:2010}, a modulation at finite ${\bf Q}$ will not show net IUC nematicity, just as antiferromagnets have no net magnetization.
However, from a strong coupling perspective that focuses on the influence of {\it local} antiferromagnetic correlations, multiple orders could naturally intertwine.

While it is to be expected on symmetry grounds that a nematic order\cite{Kivelson:1998zr} will be more robust and may coexist with disordered stripes\cite{nie2013}, microscopic studies so far have focused on either finite $\bf Q$ ordering of spin and charge stripe phenomena\cite{poilblanc:1989, machida:1989,schulz:1990,zaanen:1989,PhysRevLett.80.1272,scalapino:2012b} or ${\bf Q}=0$ IUC nematic phenomena \cite{yamase:2000c, kee:2003, metzner:2003, yamase:2005,halboth:2000, gull:2009, okamoto:2010,fischer:2011}. Moreover, since 
simple Hartree-Fock mean-field theories 
incorrectly predict insulating {period-8} stripes\cite{poilblanc:1989, machida:1989,schulz:1990,zaanen:1989},
the present theoretical understanding of the experimentally observed metallic {period-4} charge stripes at doping $x=1/8$\cite{tranquada:1995} is dependent upon 
elaborate variational numerical studies\cite{seibold:1998, ichioka:1999, normand:2001, lorenzana:2002, seibold:2004,PhysRevLett.80.1272} or the scenario of 
Coulomb-frustrated phase separation\cite{emery:1993}.
 Our goal is to capture a wide range of spin and charge ordered states, including those 
experimentally observed, in a simple microscopic model that retains the strong-coupling aspect of hole-doping that enables holes to frustrate antiferromagnetic order in the ``parent compound''. A successful demonstration of close energetic competition between diverse outcomes from the same root of local anti-ferromagnetic correlations would be an important step towards thematic understanding of the observed diverse phenomena.

Although it is well known that motion of doped holes will frustrate the antiferromagnetic background of the cuprate ``parent compounds'' \cite{Trugman1988}, much of the work on 
this issue has largely focused on the one-band Hubbard model as a minimal framework to discuss this physics.
Nevertheless, a host of experimental observations on cuprate superconductors (e.g., electron-hole doping asymmetry of the phase diagram and unusual broken-symmetry phases with spin and charge order \cite{Comin24012014,daSilvaNeto2014,Shekhter:2013uq,blackburn:2013,chang:2012,achkar:2012,PhysRevB.86.020504,ghiringhelli:2012,Wu:2011kx,okamoto:2010,lawler:2010,Daou:2010,gull:2009,Li:2008fk,PhysRevB.78.020506,PhysRevLett.100.127002,hinkov:2008} at low hole doping) 
strongly call for a description which explicitly retains the oxygen orbitals. Such multi-orbital models of the CuO$_2$ plane, like the Emery model~\cite{emery:1987}, are, however,
even less amenable to a theoretical treatment than the one-band Hubbard model.

\begin{figure}[t]
  \begin{center}
\subfigure[]{
    \includegraphics[width=0.45\textwidth]{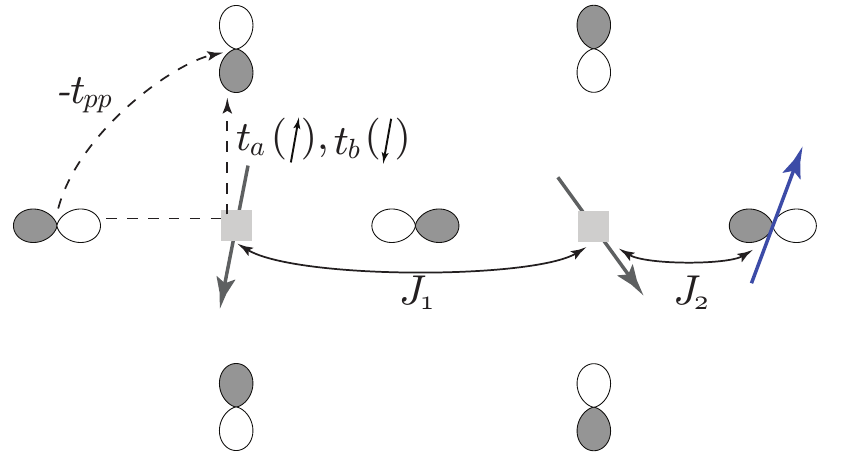}
}
\subfigure[]{
     \includegraphics[width=0.4\textwidth]{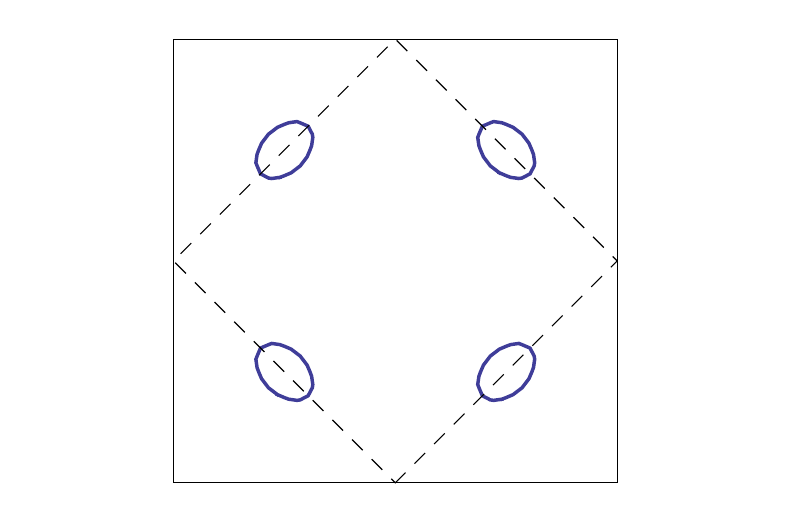}
}
\end{center}
\caption{(a) Unit cell of the CuO$_2$ plane showing a localized copper spin with mobile spinful holes on the surrounding oxygen $p_x$ and $p_y$ orbitals. 
Also indicated are the direct ($t_{pp}$) and the via-Cu hopping processes with the spin of the hole antiparallel ($t_a$) and parallel ($t_b$) to the copper spin, as well as exchange interaction parameters that define our model. 
(b) The hole-like Fermi surfaces in the antiferromagnetic N\'eel ordered state at $x\approx 5\%$ doping of our model Hamiltonian for parameter values $t_a=0.3$, $t_b=0.275$, and $t_{pp}=0.15$ (see \ref{sec:app}). This choice of parameters leads to a minimum at the nodal $(\pm \pi/2, \pm \pi/2)$ points in agreement with ARPES experiments on the extremely underdoped cuprate\cite{shen:2007b}. }
\label{fig:model}
\end{figure}

In this paper, we consider a simplified model which ignores the charge fluctuation on the transition metal sites and treat the spins on those sites as classical local moments that interact with doped itinerant holes living on the oxygen sites through a Kondo type coupling (see Figure \ref{fig:model}). This is a tractable three-orbital model that retains the spatial separation between the local moments on Cu and the doped holes that predominantly go into oxygen sites.  We consider several plausible spin-order patterns and exactly solve for the many-hole eigenstates associated with the spin-order patterns. This way we can access not only the lowest energy spin-hole configuration, 
but investigate how magnetism drives charge physics in the cuprates, study the energetic competition between a wide range of spin and charge ordered states, and gain insight 
into the interplay between spin and intra-unit-cell nematic orders\cite{lawler:2010}.

Before proceeding to the model Hamiltonian and our analysis, we clarify the nature of the charge and spin order parameters that will characterize the various phases we
consider in the paper. We define the IUC nematic order parameter, which measures the inequivalence of the hole density between the $x$- and $y$-oxygen sites within 
the CuO$_2$ unit cell as 
\begin{equation}
\eta\equiv\frac{n_x-n_y}{n_x+n_y},
\label{eq:eta}
\end{equation}
where $n_x$ and $n_y$ are the expectation value of the hole density at the $x$- and $y$- oxygen sites respectively.  The charge stripe order parameter is also extracted from the hole density as the Fourier component at a finite $\bf{Q}\neq0$, $n({\bf Q})\equiv \sum_je^{i{\bf Q}\cdot x_j}\langle c_j^\dagger c_j\rangle$. Finally, spin stripe and spiral order parameters of interest are Fourier components of the spin density at a finite $\bf{Q}$, $\langle \vec{S}({\bf Q})\rangle$ where $\vec{S}$ is either co-linear (stripe) or co-planar (spiral). 
From a symmetry perspective, any state with non-zero  $n({\bf Q})$ or $\langle \vec{S}({\bf Q})\rangle$ for one {\bf Q} (i.e., uni-directional modulation),
breaks spatial rotational symmetry as well as translational symmetry. Such a phase is referred to as electronic smectic \cite{Kivelson:1998zr}. When the modulation vector lies along the Cu-O bond direction, there will be a symmetry allowed coupling between the modulation order parameters and the IUC nematic order parameter $\eta$ resulting in $\eta\neq0$ in the presence of long range modulational order,i.e., $n({\bf Q})\neq0$ or $\langle \vec{S}({\bf Q})\rangle\neq0$.  Hence, a discussion of the IUC nematic order parameter 
for our ansatz modulational states may appear inconsequential. However, when heterogeneity and thermal fluctuations cause the experimentally observed modulational orders to
be short-ranged, a microsopic understanding of how the IUC nematic order parameter $\eta$ can locally relate to spin and charge modulations is crucial in making contact between theory and experiments \cite{2014arXiv1407.4480F}. Hence, we believe it is important to study the microscopic behavior of both order parameters even for the case of ideal long-ranged modulations.

The rest of the paper is organized as follows. In section \ref{sec:model} we introduce and motivate the model as well as the choice of some parameters. In section \ref{sec:config} we discuss the choice of spin-ordering configuration ans\"atze. In section \ref{sec:charge} we discuss the charge-ordered states we obtain associated with each spin-ordering configuration. In section \ref{sec:pd} we discuss the energy differences between different ans\"atze and the phase diagram. In section \ref{sec:Vpp} the effect of a nearest-neighbor oxygen-oxygen interaction is analyzed.
Finally, we conclude with a summary and discussion in section \ref{sec:summary}.

\section{Model}
\label{sec:model}
In this paper we restrict our attention to the Hilbert space with singly occupied Cu $d_{x^2-y^2}$ orbitals represented as local moments that interact antiferromagnetically.
The doped holes will be assumed to go into the oxygen $p_x$ 
or $p_y$ orbitals\cite{emery:1987}. These holes couple to the local moments on Cu sites through both (spin-dependent) hopping between O sites mediated by Cu sites and spin-spin interactions. Treating the moments on the Cu sites classically, we arrive at a model Hamiltonian
\begin{eqnarray}\nonumber
\!\!\!\!\!\!\!\!\!\!\!\!\!\!\!\!\!\!\!\!\!\!\!\!\!\!\! \HH&=& \frac{t_a-t_b}{2} \sum_{\langle IiI'\rangle,s}\hat p_{Is}^{\dag}\hat p^{\phantom{\dag}}_{I' s}
+ (t_a+t_b)\sum_{\langle IiI'\rangle,s,s'} \vec{S}_i\cdot(\hat p_{I s}^{\dag}\vec{\sigma}_{s s'}\hat p^{\phantom{\dag}}_{I' s'}) 
- t_{pp} \sum_{\langle I I'\rangle,s}\hat p_{Is}^{\dag}\hat p^{\phantom{\dag}}_{I' s} \\
\!\!\!\!\!\!\!\!\!\!\!\!\!\!\!\!\!\!\!\!\!\!\!\!\!\!\!
&& + J_1\sum_{\langle i,j \rangle} \vec{S}_i\cdot\vec{S}_j + J_2\sum_{\langle i, I \rangle,s,s'}\vec{S}_i\cdot (\hat p_{I s}^{\dag}\vec{\sigma}_{ss'}\hat p_{I s'}),
  \label{eq:Heff2}
\end{eqnarray}
where $\hat{p}_{Is}$ are the oxygen hole annihilation operators at oxygen site $I$ with spin $s$, $\vec{S}_i$ is a classical spin vector at Cu-site $i$ with $|\vec S_i|=1/2$, and $\vec{\sigma}$ are the Pauli matrices. 
The first two terms represent two hopping processes through the Cu sites, the $t_a$ ($t_b$) process with hole spin anti-parallel (parallel) to $\vec{S}_i$. These hopping processes amount to nearest- and next-nearest-neighbor hopping on the lattice of oxygen sites. Note that the second term in equation~\eref{eq:Heff2} 
hence introduces coupling between the motion of the holes on O-sites and the spin on the Cu sites.  The third term represents direct hopping between O-sites, which amounts to nearest-neighbor hopping on the lattice of oxygen sites. The last two terms represent exchange coupling between Cu spins (the $J_1$ term) and between a Cu spin and the hole spin on a neighboring O site (the $J_2$ term). 

The model Hamiltonian of equation~(\ref{eq:Heff2}) is a simplified form of a perturbative expansion in Cu-O hopping $t_{pd}$ (to order $t_{pd}^4$) in the large Coulomb interaction limit of the  
 Emery model Hamiltonian for the cuprates~\cite{emery:1987}:  $\HH = \HH_0 + \HH'$ with
\begin{equation}
  \HH_{0} = t_{pd}\!\!\sum_{\langle i,I\rangle,s}(\hat{d}^{\dag}_{is}\hat{p}^{\phantom{\dag}}_{I s} + {\rm h.c.})
  - t_{pp}\!\!\sum_{\langle I, I' \rangle,s}\hat{p}^{\dag}_{I s}\hat{p}^{\phantom{\dag}}_{I' s} 
  - \mu \sum_{i,s}\hat{n}^{d}_{i s} - (\mu - \Delta)\sum_{I,s}\hat{n}^{p}_{I s}
  \label{eq:hopping}
\end{equation}
and 
\begin{equation}
  \HH' = U_{d}\sum_{i}\hat{n}_{i\uparrow}^{d}\hat{n}_{i\downarrow}^{d} + U_{p}\sum_{I}\hat{n}_{I\uparrow}^{p}\hat{n}_{I\downarrow}^{p}
  + V_{pd}\!\!\sum_{\langle i, I \rangle,s, s'}\hat{n}_{i s}^{d}\hat{n}_{I s'}^{p}
  + V_{pp}\!\!\sum_{\langle I,I'\rangle,s,s'}\hat{n}_{I s}^{p}\hat{n}_{I' s'}^{p}.
  \label{eq:interaction}
\end{equation}
However, the actual model obtained through such perturbative expansion is still highly non-trivial as it does not allow double occupancy at oxygen sites and the spins on Cu-sites should be quantum mechanical spins. Frenkel et al\cite{frenkel:1990} studied the exact quantum ground state of such a model  in the presence of a single  hole  in a small cluster. Despite the reduction of the Hilbert space due to the constraint of singly occupied Cu  $d_{x^2-y^2}$ orbitals, the largest cluster they could study was a $4\times 4$ Cu-O cluster which would be too small to see the observed stripe phenomena.  
Such studies have been extended to larger
clusters \cite{PhysRevLett.106.036401} with the aim of understanding spin 
polaron formation in a 3-band model. 
However, this exact diagonalization study is restricted to 1-hole and 2-hole states in the undoped insulator and thus cannot address doping
dependent competing orders.
By treating the 
local moments on Cu sites to be classical and relaxing the no-double-occupancy constraint on oxygen sites, we arrive at the more tractable
model Hamiltonian of equation~(\ref{eq:Heff2}).  

Our simple model has several features. Firstly, it is a solvable model that retains much of the microscopic details and non-trivial interactions of under-doped cuprates. The solvability of the model allows us to study a wide variety of states including incommensurate orderings.
Secondly, we expect similar effective models are applicable to
other doped strongly-correlated charge-transfer insulators such as the nickelates. In particular, a classical approximation of the local moment on the transition-metal ion
is likely to be a better approximation in Ni given its larger moment.
Finally, through spatial separation between spins and holes, the model offers a rich playground for strong-coupling-driven spin and charge orders co-existing with  intra-unit-cell nematic order.

Earlier consideration of the strong-coupling limit of the Emery model by Kivelson et al.~\cite{Kivelson2004} focused on the limit of vanishing inter-oxygen site hopping. In that limit, the dynamics is strictly one-dimensional and hence the ground state is a nematic phase. Three key differences between the limit considered in Ref.~\cite{Kivelson2004} and the limit captured by our model equation~\eref{eq:Heff2} are that 1) we have taken $U_d$ to be so strong to the extent that we suppressed  the charge fluctuation in the Cu sites, 2) we consider inter-oxygen hopping to be comparable to exchange interactions, and 3) we take the Cu-O exchange interaction into account. As a result, the charge dynamics in our model is strictly two dimensional. However we make the approximation of leaving out $U_p$ and $V_{pp}$ which yields an exactly solvable model for the hole motion on a system size that can exhibit translational symmetry breaking. This approximation may not affect the conclusions qualitatively in the limit of dilute hole density.

Although the model Hamiltonian has quite a few parameters, many of them are constrained by experiments.
We use a Cu-Cu exchange interaction $J_1 \sim 125$meV, close to the value estimated for La$_2$CuO$_4$ from neutron scattering studies of
the spin-wave dispersion \cite{Hayden1991}.
In order to constrain hopping strengths, we examine 
the dispersion of a single
hole doped in an ordered N\'eel antiferromagnet. 
We can diagonalize the hole Hamiltonian assuming an out of plane spin moment on Cu atoms in an antiferromagnetic arrangement, {\it i.e.} $S_i^z=\frac{1}{2}(-1)^{i_x+i_y}$, 
and obtain four dispersing bands (see \ref{sec:app}). 
The resulting lowest dispersion can be compared to 
the ARPES measurements on extremely underdoped cuprates \cite{shen:2007b}. Although the ARPES energy distribution curves are broad, we can infer the following
energy scales from the peak positions.
The binding energy at $(\pi/2,\pi/2)$ is lower than that at $(\pi,0)$ by about $300$meV\cite{shen:2007b}, which corresponds to $2 t_{pp}$; this fixes $t_{pp}=150$meV.
Shen et al.~\cite{shen:2007b} also found that the spectral peaks disperse by about $1.4$eV going from the $\Gamma$-point
to $(\pi/2,\pi/2)$. In our model, this energy difference is just $4 t_b + 2 t_{pp}$; this fixes $t_b=275$meV. 
Cluster diagonalization calculations in Ref.~\cite{frenkel:1990,Nazarenko1995} indicate
that $t_a$ is the same sign as $t_b$, with $|t_b-t_a| \ll t_a+t_b$. We therefore examine a range of values $t_b/2 < t_a < 3 t_b/2$.
The above choice of hopping parameters yields a Fermi surface consisting of
hole-pockets centered at $(\pm\pi/2,\pm\pi/2)$ for the $\hat p_{Is}$ holes (see Fig. \ref{fig:model}(b)), consistent with the energy minimum of a single hole
in the N\'eel ordered antiferromagnet being centered at $(\pm\pi/2,\pm\pi/2)$. We expect the antiferromagnetic Cu-O exchange interaction  $J_2$, which is 
only present at finite doping, to be stronger than the Cu-Cu exchange $J_1$. Hence we explore a range of $J_1< J_2<3J_1$. 
We note that our values for the parameters
in the effective Hamiltonian for doped holes differ slightly from those used in previous work \cite{frenkel:1990,Nazarenko1995,Starykh1995} - 
the values we use should be viewed as effective couplings given our assumption of a classical copper spin.
They also place the regime of interest in the strong coupling limit, since $t_a+t_b$ and $J_2$ are much greater than the free fermion bandwidth set by 
$|t_a-t_b|\sim t_{pp}$ and motivate the use of a variational approach as discussed in the next and following sections. 

\section{Ans\"atze for spin-order configurations}
\label{sec:config}
The key mechanism by which the model Hamiltonian of equation~(\ref{eq:Heff2}) drives spin and charge order at finite doping is the frustration of antiferromagnetic order in the parent compound (i.e., zero doping). 
The undoped cuprate and nickelate insulators are ordered antiferromagnets, possessing long-range
N\'eel order in planes which are stacked along the $c$ axis. The planar N\'eel order has a wavevector $(\pi,\pi)$ and is a collinear state
with $\vec S_i = S \hat{n} (-1)^{i_x+i_y}$ with $\hat{n}$ being a unit vector. 
Doped oxygen holes frustrate the antiferromagnetic alignment of Cu moments driven by the Cu-Cu exchange $J_1$ 
through a strong antiferromagnetic Cu-O exchange interaction
$J_2$.
Thus, doped holes promote a ferromagnetic arrangement of the two neighboring local Cu-moments and drive spatial symmetry breaking in the spin configuration. 
Interestingly, such frustration is not expected in the
electron-doped cuprates, where doped electrons go onto Cu sites. 
This naturally explains why the N\'eel order is present to much higher doping in electron doped cuprates whereas the hole-doped cuprates exhibit various broken-symmetry states. 

While ultimately the model equation~(\ref{eq:Heff2}) can be exactly solved by combining Monte Carlo simulation of the spin problem with the exact diagonalization of the fermion problem, we here consider two classes of spin-order configurations and solve the fermion problem within each class. This approach offers us an understanding of the parameter-space landscape. Moreover, 
although our understanding is limited to the set of ansatz spin patterns we consider, this approach has the advantage over the explicit solution in that it allows us to compare energetics of different candidate states. The two classes we consider are coplanar spin spirals and collinear spin stripes with different wave vector $\bf Q$. As there is much literature on both of these candidate states we defer in-depth discussion to review articles (see e.g., \cite{Vojta2009,Kivelson2003}). Rather, we discuss aspects of these states that are directly relevant to our study and the rationale for their consideration below.

(i) {\bf Coplanar spin spirals:} While a doped hole promotes ferromagnetic correlations of the local moments in its vicinity, the spins far from the hole maintain their  N\'eel correlations. Hence, in the absence of magnetic anisotropy a coplanar spiral state may be expected as a solution that interpolates between the N\'eel order and ferromagnetism when holes are doped into a quantum antiferromagnet as it has been argued in Ref.~\cite{PhysRevB.42.2485}.
However, spiral states are accompanied by translationally invariant charge distribution and they are unstable against magnetic anisotropy as well as  phase separation. Further, susceptibility studies\cite{PhysRevB.78.214507} in spin-ordered cuprates are consistent with collinear order.
However, the smoking-gun polarized neutron scattering experiment has not been performed to date and local spiral distortions of the spin background may be an appropriate picture in the insulating regime at low doping
\cite{PhysRevB.73.085122}.

\begin{figure}
\includegraphics[width=.3\textwidth]{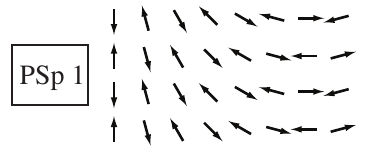}
\includegraphics[width=.3\textwidth]{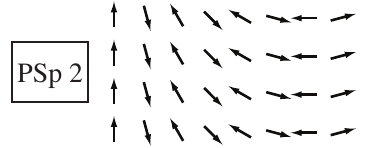}
\includegraphics[width=.3\textwidth]{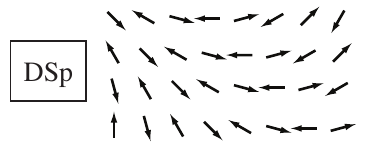} \\
\includegraphics[width=.3\textwidth]{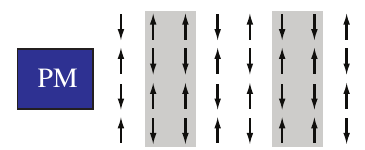}
\includegraphics[width=.3\textwidth]{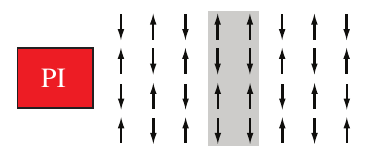}\\
\includegraphics[width=.3\textwidth]{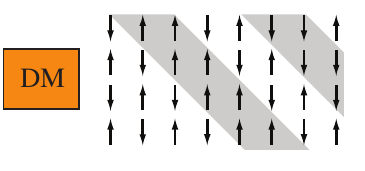}
\includegraphics[width=.3\textwidth]{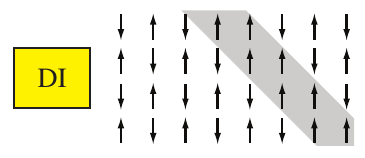}\\
\caption{Catelog of Ans\"atze used in our variational calculations in the form of the classical spin configurations of local moments on Cu atoms. The gray areas in the stripe patterns denote anti-phase domain walls between antiferromagnetic domains.}
\label{fig:patterns}
\end{figure}

Below, we pick the spiral to lie in the $S_x-S_y$ plane, setting $\vec S_i = S (\cos {\bf Q} \cdot {\bf r}_i, \sin {\bf Q} \cdot {\bf r}_i,0)$, and
consider different propagation wavevectors ${\bf Q}$ running parallel to the Cu-O bonds (labelled {\bf Psp1}, {\bf Psp2}) or along
the diagonal direction ({\bf Dsp}), see Fig.~\ref{fig:patterns}, and compute their physical properties and energies.  
We
show that within a three-orbital picture, the spin spirals with a wavevector parallel to the Cu-O bonds are naturally accompanied by a translationally invariant oxygen 
hole density, but with {\it charge nematic order} on the oxygen sites as has been observed in STS experiments\cite{lawler:2010}. 
This opens up a possibility of disordered spiral leaving the discrete and robust nematic order as the only observable effect. Similar ideas have been perviously
considered in the context of frustrated magnets \cite{PhysRevB.81.214419}.

(ii) {\bf Stripes:} 
Two theoretical approaches discussed stripes prior to their experimental observations. The first approach was that of Hartree-Fock mean-field theory
\cite{poilblanc:1989, machida:1989,schulz:1990,zaanen:1989} which predicted insulating (fully filled) { period-8} charge stripes at $x=1/8$ doping. 
The other approach was built on the observation that phase separation should be expected in $t$-$J$ models for $J \gg t$, where `slow'
doped holes cluster together to avoid disrupting the N\'eel order of the local moments\cite{PhysRevLett.64.475}. 
This
led to the proposal that stripe orders could emerge as periodic density modulations due to frustration of such phase separation by long-range Coulomb
interactions\cite{emery:1993}. 
Following the experimental observation of metallic {period-4} charge stripe at $x=1/8$ doping in La-based cuprates\cite{tranquada:1995}, 
theoretical efforts focused on going beyond Hartree-Fock mean-field theory to obtain the observed metallic stripe \cite{seibold:1998, ichioka:1999, normand:2001, lorenzana:2002, seibold:2004,PhysRevLett.80.1272}. A particularly convincing case was made by a density matrix renormalization group study by White and Scalapino\cite{PhysRevLett.80.1272} which not only obtained the observed metallic stripe ground states at $x=1/8$ but also 
demonstrated that anti-phase domain walls between collinear antiferromagnetic domains attract
charge stripes.
This issue of charge stripe periodicity (period-4 and metallic v.s. {period-8} and insulating) bears particular importance in the discussion of the role of charge stripes in superconductivity\cite{scalapino:2012b}. However, most previous efforts at going beyond Hartree-Fock mean-field theory and incorporating strong-coupling physics relied on sophisticated numerics. The spatial separation of Cu-moment and doped holes in our model allows us to investigate the complex spin-charge interplay in the strong-coupling regime in a transparent manner. 
In our model, spin-antiphase domain walls can form upon hole-doping to relieve the frustration between 
the AFM order favored by 
Cu-Cu antiferromagnetic superexchange interactions and the ferromagnetic order favored by an antiferromagnetic Cu-O exchange. 

We model the spin stripes as antiphase domain walls in the collinear antiferromagnet and consider several different uni-directional domain-wall configurations depending on the direction of periodicity as well as the period. Note that these spin stripes will be bond-centered by construction. We consider parallel stripes with wave vectors ${\bf Q}$ along the Cu-O bond directions as well as diagonal stripes with ${\bf Q}$ at 45$^o$ angle with respect to Cu-O bond direction. At fixed hole density of $x=1/8$, different periods determine whether the stripe can support metallic transport along the stripe: period-4 parallel charge stripe (labeled ${\bf PM}$, see Fig 2) will be half-filled and metallic while period-8 parallel charge stripe (labeled ${\bf PI}$, see Fig 2) will be fully-filled and insulating. Similarly, one can consider metallic (labeled ${\bf DM}$, see Fig 2) and insulating (labeled ${\bf DI}$, see Fig 2) diagonal stripes.

\section{Charge orders associated with different spin-order ans\"atze}
\label{sec:charge} 
For each spin-order candidate, we diagonalize the (quadratic) hole Hamiltonian and use the resulting lowest-energy many-hole wave function and the spin configuration to evaluate the total energy for the spin-charge-ordered state. The total energy for a given spin configuration $\{{\vec{S}}\}$ at a given doping $x = n/N$, with $N$ the number of Cu-O unit cells, is given by
\begin{equation}
  E(\{\bS\})/N = \frac1N\sum_{l<n}\xi_l(\{\bS\}) + \frac1N E^{\rm Cu-Cu}(\{\bS\}),
  \label{eq:totalenergy}
\end{equation}
where $\xi_l(\{\bS\})$ are the (energy-ordered) eigenvalues of the Hamiltonian \eref{eq:Heff2} for the given configuration $\{\bS\}$. In equation~\eref{eq:totalenergy} the last term is the interaction energy due to Cu-Cu exchange (the $J_1$ term in equation \eref{eq:Heff2}).

\begin{figure}
 \includegraphics[width=\textwidth]{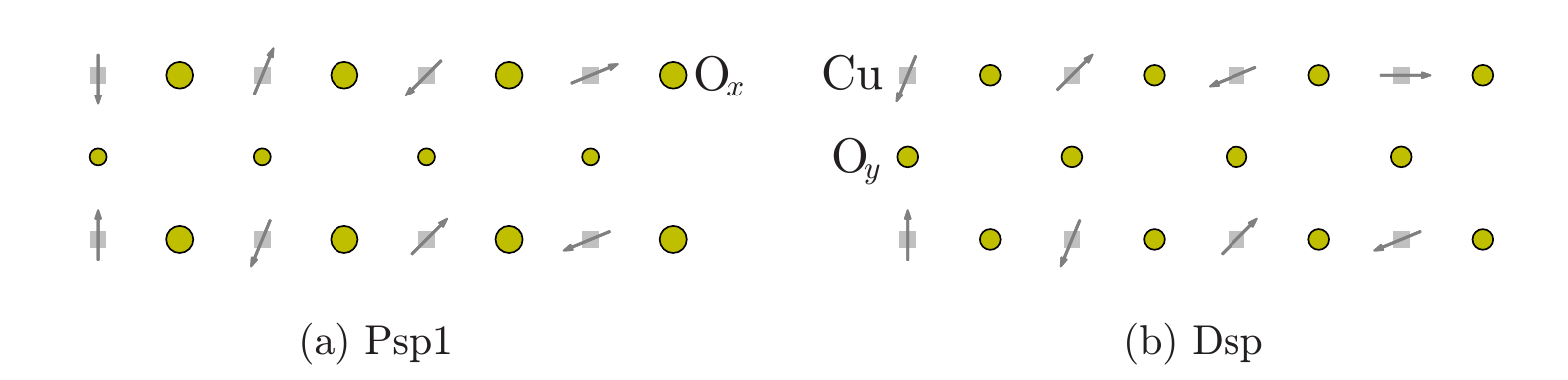}
\caption{A representative oxygen-hole charge distribution $n_{x,y}$  are shown through yellow circles for  (a) a parallel spiral Psp1 and (b) a diagonal spiral with $x = 1/8$. The size of yellow circles represent the magnitude of charge density at the site. The parallel spiral shows non-zero nematic order $\eta\neq0$.
}\label{fig:charge-sp}
\end{figure}

For the spiral configurations given by $\bS_i=S(\cos \bQ\cdot{\bf r}_i, \sin \bQ\cdot{\bf r}_i, 0)$ we take advantage of a closed form of the Hamiltonian equation~\eref{eq:Heff2} in  momentum space (See \ref{sec:app}) to find the wave vector $\bQ$ that minimizes the energy for each spiral. Fig.~\ref{fig:charge-sp} shows the lowest-energy charge distribution for a parallel spiral (Fig.~\ref{fig:charge-sp}(a)) and a diagonal spiral (Fig.~\ref{fig:charge-sp}(b)) spin configurations. Though it is well known that spiral order of spins would not be accompanied by any charge modulation, Fig.~\ref{fig:charge-sp}(a) shows that parallel-spiral tendency drives IUC nematic charge order (for $t_a=200$meV and $J_2=200$meV, we find $\eta\approx 5\%$ for the Cu-spin configuration in Figure~\ref{fig:charge-sp}). 
Note that even when the spiral order of the spins is disordered due to thermal or quantum fluctuations, the discrete symmetry breaking of IUC nematic in the charge sector can be more robust.

\begin{figure}
 \includegraphics[width=0.85\textwidth]{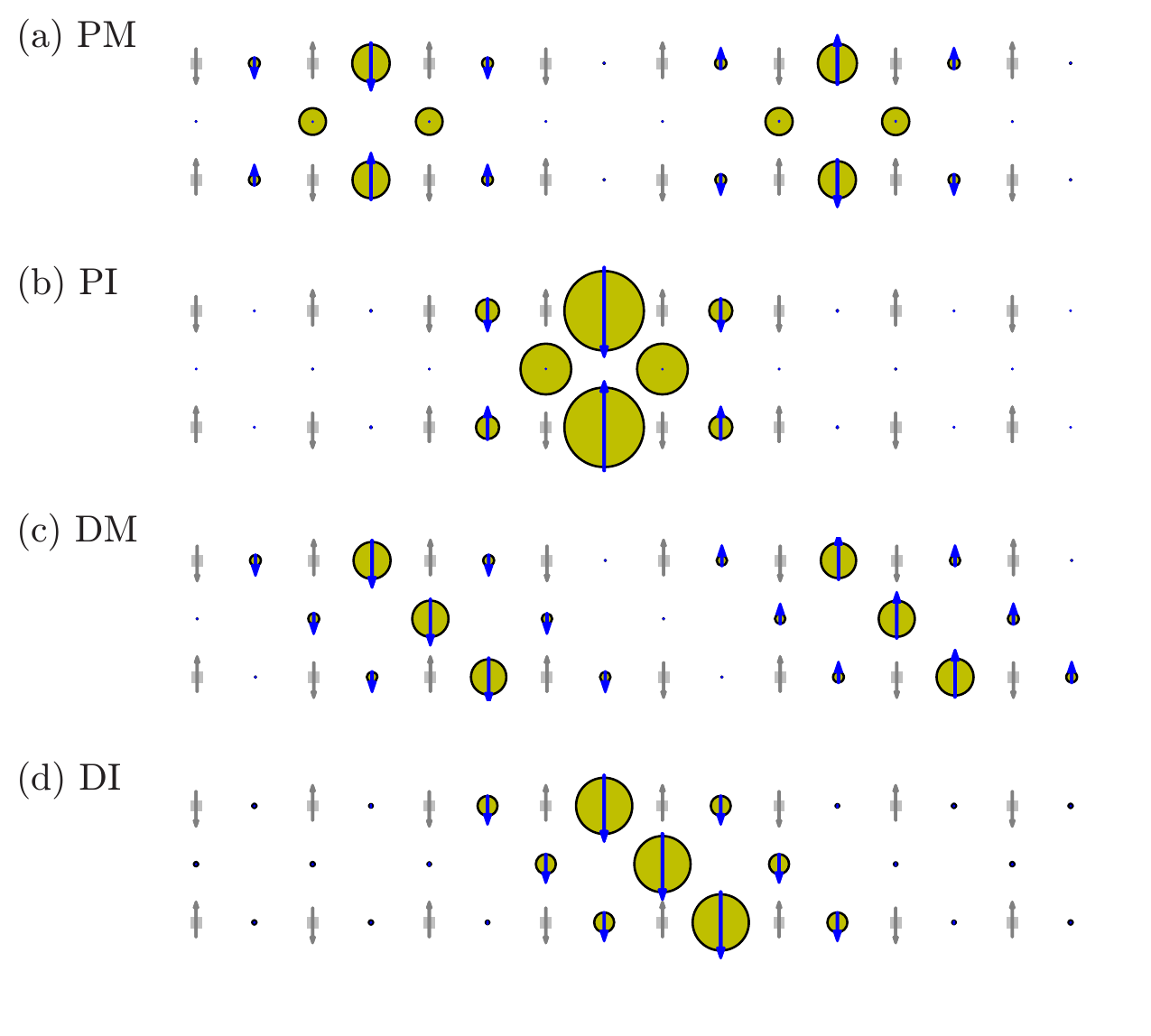}
\caption{Charge distribution (depicted by the size of the yellow circles) and spin polarization (length of blue arrows) of oxygen holes for (a) metallic and (b) insulating parallel stripes, as well as (c) metallic and (d) insulating diagonal stripes for $t_a=200$meV and $J_2=200$meV. Note that the length of the (spin) arrow on the oxygen site is scaled by a factor of 5 as compared to the Cu sites. }
\label{fig:oxygen}
\end{figure}

For the collinear stripe configurations, we consider a lattice of $32\times32$ unit cells (2048 oxygen sites) and diagonalize the quadratic Hamiltonian of the holes only living on the oxygen site, equation~\eref{eq:Heff2}. The resulting charge-order patterns for different trial spin configurations (see Fig.~\ref{fig:oxygen}) clearly show that the holes are attracted to the anti-phase domain walls in the spin configurations driven by the Cu-O exchange-coupling $J_2$.  The kinetic terms broaden the hole distribution and favor the metallic charge stripe. It is remarkable that our simple model can readily access the spin and charge striped ground-state configuration reminiscent of those obtained in density matrix renormalization group studies of the $1/8$-doped $t$-$J$ model\cite{PhysRevLett.80.1272}. Moreover, as this model incorporates the mostly-oxygen character of doped holes, the charge stripes centered at the anti-phase domain wall of the antiferromagnetic background are naturally coupled to IUC nematic. The parallel stripe configurations obtained in Fig.~\ref{fig:oxygen} clearly demonstrate a coupling between the (Ising) IUC nematic order and the stripe-ordering wave vector. In the sense of Ginzburg-Landau theory of order parameters,  
Fig.~\ref{fig:oxygen} demonstrates a coupling between the Ising nematic order parameter and the difference in amplitude of the charge-density wave (CDW) order parameters for CDW's propagating along the two Cu-O bond directions\cite{Mesaros2011} at a microscopic level. The charge distribution we obtain for trial spin configurations in Figs.~\ref{fig:charge-sp} and \ref{fig:oxygen} makes it clear that any spin order with modulation vector along the Cu-O bond direction will be accompanied by IUC nematic irrespective of charge stripe order. \footnote{On symmetry grounds, any unidirectional spin order along Cu-O bond direction breaks the point group symmetry and hence it can, in
principle, couple to nematic order parameter; what is new here is an explicit microscopic realization of such a coupling.} The remaining question is which of these candidate states is lowest in energy and what are the energy differences between competing states. 

\section{Phase diagram}
\label{sec:pd}
\begin{figure}[th]
  \centering
\includegraphics[width=.45\textwidth]{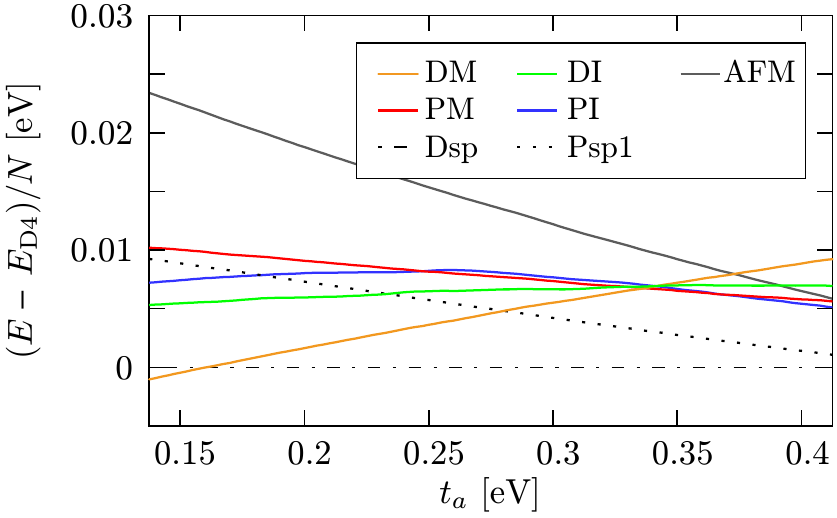}
  \caption{The energy difference of all the states in our Ansatz catelog (see Fig. \ref{fig:patterns}) compared to the diagonal spiral state (Dsp) as a function of $t_a$ for  $J_2=170$meV (dashed line in the phase diagram in Figure~\ref{fig:pd}(b)). Shown are the antiferromagnetic (AFM), the parallel spiral (Psp1), the diagonal spiral (Dsp), as well as the diagonal metallic stripe (DM), diagonal insulating stripe (DI) parallel metallic stripe (PM), and parallel insulating stripe (PI). Notice the parallel orders are preferred for larger $t_a$.   }
  \label{fig:de}
\end{figure}

While our variational study is limited by the choice of states we consider, it has the advantage of allowing the energetic comparison between different candidate states over Hartree-Fock or variational Monte Carlo studies. Surprisingly, we find the entire collection of states we consider to show close energetic competition. For the most part of the parameter space we consider (varying hopping through Cu $t_a$ and Cu-O exchange $J_2$) different states differ in energy only by a few meV. 
The close energetic competition is clear in
 Figure~\ref{fig:de} which shows energy differences as a function of $t_a$ at fixed $J_2 = 170$meV. Such close competition indicates small perturbations to our model Hamiltonian such as spin-orbit coupling or lattice anisotropies could change the phase diagram significantly.

\begin{figure}[ht]
  \centering
\subfigure[]{
\includegraphics[width=.45\textwidth]{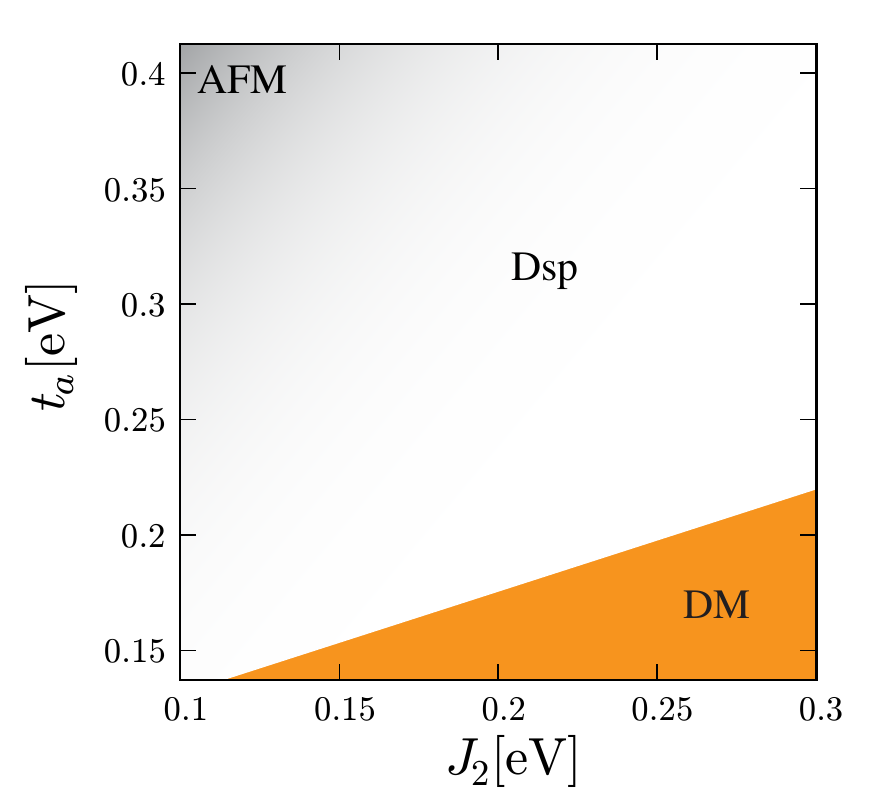}
}
\subfigure[]{
\includegraphics[width=.45\textwidth]{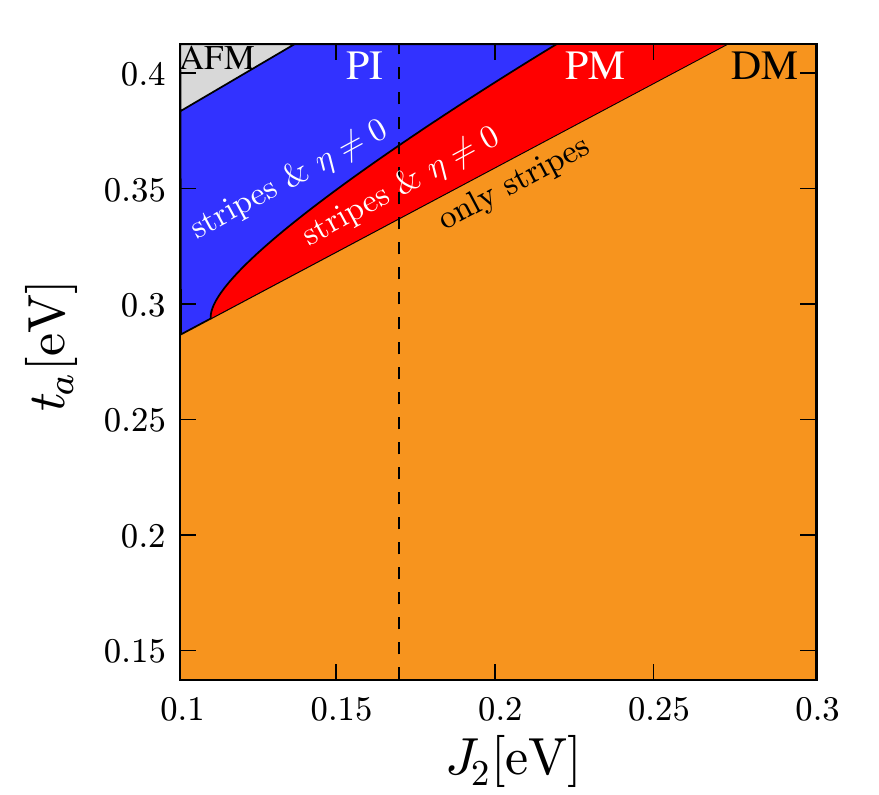}
}
\caption {(a) Full phase diagram at $x=1/8$ with all phases allowed as a function of the Cu-O spin-spin interaction $J_2$ and the hopping $t_a$. (b) the phase diagram if we only allow collinear order for the Cu spins. The parameters used are $t_b = 275$meV, $J_1 = 125$meV and $t_{pp}=150$meV. Parallel stripes show non-zero IUC nematicity $\eta\neq0$ (see Eq.~\ref{eq:eta}).}
  \label{fig:pd}
\end{figure} 

It is well known that incorporating quantum fluctuations of spins strongly favors collinear order over coplanar or non-coplanar states \cite{Henley1989,Galitski2014},
Similarly, spin anisotropy due to weak spin-orbit coupling could also hinder coplanar spirals. We therefore present two phase diagrams here: a 
phase diagram that includes co-planar spiral and a  phase diagram that only concerns collinear orders. 
Figure~\ref{fig:pd}(a) shows the phase diagram that allows for co-planar spiral
as a function of hopping through Cu-sites $t_a$ and the Cu-O exchange $J_2$.
 We have limited the plotting range of $J_2$ to be a reasonable range of $100$meV $<J_2<300$meV. Figure~\ref{fig:pd}(a) shows that the lowest-energy configuration with the parameters motivated from cuprates is dominated by diagonal orders such as diagonal spiral or diagonal metallic stripe. 
In the limit of large $J_2$ (not shown), parallel spiral with IUC nematic appears before Cu spins align ferromagnetically.
 Figure~\ref{fig:pd}(b) shows the phase diagram restricted to collinear orders. Notably metallic diagonal and parallel stripe orders appear in a substantial region of the phase diagram. The close competition between diagonal stripe order and parallel stripe order is remarkable in light of the experimentally known transition from diagonal to parallel stripe upon doping\cite{PhysRevB.70.104517}. The parallel metallic charge stripe order (PM) at $x=1/8$ is consistent with observations in La-based compounds\cite{PhysRevB.70.104517,PhysRevLett.85.1738} as well as those in Bi-based compounds \cite{lawler:2010,Mesaros2011,
Comin24012014,daSilvaNeto2014}. 

The microscopic tie between spin-ordering patterns and charge-ordering patterns resulting from our simple model depicted in Figs.~\ref{fig:charge-sp} and \ref{fig:oxygen} hints at a microscopic mechanism of charge order without spin order at finite temperature\cite{Sun2010}. The spin orders under consideration require breaking
of spin rotational symmetry as well as spatial translation and point-group symmetries while charge orders only break spatial symmetries. 
While the driving force for charge order in our model is the frustration of antiferromagnetic order upon holes entering oxygen sites, 
fluctuations in spin space will suppress spin order.
 It has been shown in the context of Fe-based superconductors\cite{PhysRevB.85.024534} that spin fluctuations $\langle| {\bS}_{\bf Q}|^2\rangle\neq0$ in the absence of 
spin order ($\langle {\bS}_{\bf Q}\rangle=0$) can drive nematicity. Similarly, since the cuprates are quasi-2D systems, it is quite plausible that
thermal fluctuations may prevent spin order in the stripe state, while $\langle| {\bS}_{\bf Q}|^2\rangle\neq0$ may still leave the charge order visible. 
This can be easily seen by the fact that the charge stripe and IUC nematic in the parallel stripe phases in Fig.~\ref{fig:oxygen} as well as IUC nematic in the parallel spiral phase in Fig.~\ref{fig:charge-sp} are insensitive to the spin-orientation, so (thermal) averaging over the global spin orientation will leave these orders unaffected. 
Hence, charge order in Fig.~\ref{fig:charge-sp} can onset without detectable spin order or with a lower temperature
transition into a state with coexisting spin order. This would amount to a microscopic realization of a so-called ``charge-driven transition'' in the Landau theory study Ref.~\cite{PhysRevB.57.1422}.
Experimental observations of charge and/or spin order in cuprate families broadly show such preferential visibility of charge order \cite{PhysRevB.70.104517,PhysRevLett.85.1738,lawler:2010,Mesaros2011,achkar:2012,chang:2012,blackburn:2013,Comin24012014,daSilvaNeto2014}.

\section{Effects of inter-oxygen repulsion $V_{pp}$}
\label{sec:Vpp}
As IUC nematic naturally accompanies all modulations along the Cu-O bond directions within our model, it is natural to ask what would be the effect of the inter-oxygen repulsion  $V_{pp}$ that was found to drive IUC nematic~\cite{fischer:2011,Kivelson2004}. Recently Bulut et al.~\cite{PhysRevB.88.155132} extended the mean-field study by two of us~\cite{fischer:2011} to include the charge-density wave instability as well as IUC nematic instability. They found the d-form factor component $\psi^d(\bvr)$, defined on oxygen sites using the notation of Ref.~\cite{PhysRevLett.111.027202} via
\begin{equation}
\rho(\bvr) = {\rm Re} \left[ \left( f^s(\bvr) \psi^s(\bvr) + f^{s'}(\bvr) \psi^{s'}(\bvr) + f^d(\bvr) \psi^d(\bvr) \right) e^{i\bQ \cdot \bvr}\right],
\end{equation}
where $f^{s}(\bvr)$ ($f^{s'}(\bvr)$) is zero (one) on copper sites and one (zero) on oxygen sites and $f^{d}(\bvr)$ is zero on copper sites, one on x oriented oxygen sites and minus one on y-oriented oxygen sites, to dominate the density wave instability along the Brillouin zone diagonal.
While different form-factor components of density waves are not symmetry distinct, experimental observation of the d-form factor \cite{Comin24012014,2014arXiv1404.0362F} hints at the importance of microscopic interactions promoting an anti-phase relation between the two oxygen sites\cite{2014arXiv1404.0362F}. 

As charge fluctuations on Cu-sites have been projected out in our model ($\psi^s(\bvr)=0$), charge stripes found in our model consist only of s'- and d-form factor components. In order to study the effect of $V_{pp}$ on the charge configurations associated with candidate spin configurations, we treat the $V_{pp}$ term at the level of self-consistent Hartree approximation. As experiments observe $\bf{Q}=0$ IUC nematic simultaneously with short-ranged d-form factor density waves~\cite{2014arXiv1404.0362F} we focused the study to the parameter space where a Cu-O bond-direction charge stripe is present in the absence of $V_{pp}$. 

\begin{figure}
\centering
\includegraphics[width=.5\textwidth]{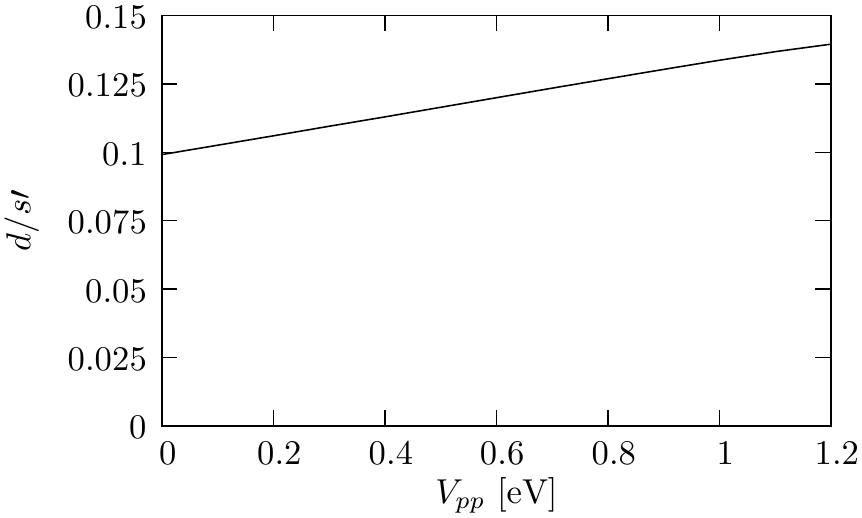}
\caption{The effect of $V_{pp}$ on the relative importance between s'-form factor and d-form factor components of charge-density waves for $t_a=350$meV and $J_2=170$meV for the PM state.}
\label{fig:dff}
\end{figure}

On symmetry grounds, $\psi^{s'}(\bvr)$, $\psi^d(\bvr)$ and IUC nematic order parameter can form a cubic coupling within Landau theory. Hence, we generically expect both s'- and d- form factor components to be present given the robust IUC nematicity coexisting with charge stripes in our charge-ordering patterns. For $t_a=350$meV, $J_2=170$meV, 
figure \ref{fig:dff} shows the resulting ratio of the d- and s'- form factor components. While the density waves we obtained for the PM configuration have predominantly s'-form factor, Fig.~\ref{fig:dff} clearly shows that $V_{pp}$ promotes a d-form factor for the charge stripe along the Cu-O bond direction with the wave vector consistent with the experimental observations.

\section{Discussion}
\label{sec:summary}
To summarize, we considered a three-orbital model for underdoped cuprates, which incorporates strong-coupling physics through singly occupied Cu-sites hosting local moments and reflects the charge-transfer energy through constraining holes to live on the oxygen sites. 
We took a variational approach for the spin configurations and solved for many-hole states exactly for each configuration at $x=1/8$ doping.
We found that the balance between the Cu-Cu exchange interaction, the kinetic energy of holes and the Kondo type coupling between the Cu spins and the spin of the holes conspire to a rich phase diagram featuring several experimentally observed phases. What is more interesting is the close energetic competition between the candidate states such as co-planar spirals, diagonal stripes and parallel stripes with different wave lengths. Within this model, any parallel order including co-planar spiral orders exhibit non-zero IUC nematic order parameter, providing a microscopic mechanism for robustness of a  IUC nematic.
The mechanism for nematicity considered in this paper differs from the weak-coupling fermi-surface instability driven by a so-called $F_2$ interaction \cite{metzner:2003,yamase:2005} or inter-oxygen repulsion $V_{pp}$\cite{fischer:2011} in that we focus on the strong
local antiferromagnetic correlations between singly occupied Cu sites as the driving force for nematicity.

Though we studied the parameter space constrained by various experiments on the cuprates, our model can be applied to 
other transition metal oxides like the doped nickelates which exhibit similar spin-charge ordered states as cuprates~\cite{PhysRevLett.73.1003}. In the nickelates, a local moment at Ni sites may arise from the effect of strong Hund's coupling between two electrons 
(or two holes) in the two $e_g$ orbitals, leading to an orbital singlet and a spin-triplet $S=1$ state. This spin-triplet nature of the local moment suppresses hopping through Ni sites which renders the system insulating at all doping ($x<0.9$) without ever exhibiting superconductivity. Antiferromagnetism is more robust in Nickelates and extends to hole concentration of $n_h=x+2\delta\sim 0.2$ before it is replaced by fully filled diagonal stripe order\cite{PhysRevLett.73.1003}. Indeed, we find the diagonal insulating stripe as most favored co-linear order in the limit $t_a\ll J_1< J_2$.

Recently, several model studies investigating the Fermi-surface instability towards formation of charge-density waves\cite{Efetov:2013fk,PhysRevLett.111.027202, PhysRevB.85.165130,PhysRevB.86.085113,PhysRevB.88.155132,2014arXiv1404.1335A} have found charge-density waves with  predominantly d-form factor. Some theories focused on the ``hot spot'' regions of the Fermi surface, where antiferromagnetic scattering will be especially strong near the antiferromagnetic quantum critical point \cite{Efetov:2013fk,PhysRevLett.111.027202}, while others considered one-band models with d-form factor interactions\cite{PhysRevB.85.165130,PhysRevB.86.085113} or a three-band model\cite{PhysRevB.88.155132}. Though the prominence of the d-form factor component in the density waves appear compatible with experiments\cite{Comin24012014,2014arXiv1404.0362F}, these models universally obtained  charge-density waves with the wave vector along the Brillouin zone diagonal, i.e., diagonal stripes. This direction and the magnitude of the wave vector is at odds with experimental findings of parallel stripes. 
More recently, Atkinson et al~\cite{2014arXiv1404.1335A} showed that the wave vector found in Ref.\cite{PhysRevB.88.155132} can be rotated to lie in the Cu-O bond direction 
by assuming large staggered moments which reconstruct the Fermi surface. 
However, experimentally it is known that antiferromagnetic fluctuations at $(\pi,\pi)$ are replaced by incommensurate magnetic peaks at temperatures well above the charge ordering temperature in inelastic neutron scattering at low energies~\cite{Aeppli21111997,PhysRevB.70.104517}. Moreover, nematic ordering which appears to be most robust in experiments, only appear as sub-dominant instability in these Fermi-surface instability studies
\cite{Efetov:2013fk,PhysRevLett.111.027202, PhysRevB.85.165130,PhysRevB.86.085113,PhysRevB.88.155132,2014arXiv1404.1335A}.

In this paper, we took a strong-coupling approach of projecting out charge fluctuation at the Cu-sites yet dealing with an exactly solvable model by ignoring quantum fluctuations of spins at the Cu-sites. In this approach, we found ${\bf Q}=0$ IUC nematic to naturally coexist with any modulation along the Cu-O direction be it co-planar spiral or collinear spin stripes. This is in contrast with weak-coupling Fermi-surface-instability approaches finding competition between ${\bf Q}=0$ order and ${\bf Q}\neq0$ density waves. Also, we find various experimentally observed charge-ordered states (diagonal insulating stripe, diagonal metallic stripe, parallel metallic stripe) to dominate at different parts of the physically-motivated parameter space exhibiting a close competition.  The understanding of the parameter space of our model we gained from the present variational approach can guide us in the attempt for a more explicit exact solution of the model combining Monte Carlo simulations with exact diagonalization. 

Before closing, we turn to the implications of our results for the experimentally observed broken symmetry states in different cuprate families (summarized in Refs. \cite{PhysRevLett.110.017004,2014arXiv1407.4480F,2014arXiv1406.1595B}). Although d-wave superconductivity is robust in all the cuprate families, 
the recent detection of charge order in YBa$_2$Cu$_3$O$_{6+x}$ reveals material specific differences regarding the issue of competing broken symmetry states. 
The La-214 family exhibits spin order at wave vector ${\bf Q}_{s}={\bf Q}_c/2$ where ${\bf Q}_{s}$ and ${\bf Q}_c$ are spin and charge modulation wave vectors respectively,
while YBa$_2$Cu$_3$O$_{6+x}$ with $x>0.8$ does not appear to have static spin order. BSCCO and La-214 families appear to have period-4 charge
order at 1/8 doping, while YBa$_2$Cu$_3$O$_{6+x}$ has incommensurate order with a period close to $3 a_0$. In BSCCO and La-214, the charge order wavevector
grows with increasing hole doping while it appears to shrink in YBa$_2$Cu$_3$O$_{6+x}$. Nevertheless, all cuprate families indicate incommensurate spin correlations 
(static or dynamic) whose wavevector grows with doping. Can such a diverse phenomena of competing orders be captured within a unified microscopic framework?
 
Regarding the issue of spin order, our variational study shows that within our strong coupling mechanism, a long-ranged spin stripe order with wave vector ${\bf Q}_{s}$ leads to a charge stripe at wave vector  $2{\bf Q}_{s}$. On the other hand, the present study cannot address whether a charge stripe driven by local antiferromagnetic correlations would necessarily require spin order; indeed, broken lattice symmetries might be more stable against quantum and thermal fluctuations, and thus survive even if long-range spin order itself melts. For instance, it has already been shown that when charge order has a higher transition temperature, spin order may or may not occur at the level of Landau theory of coupled order parameters \cite{PhysRevB.57.1422}. Regarding the issue of the doping dependence of the ordering wave vector, our present study has only focused on one doping of $x=1/8$. Here 
we found the period-4 metallic to be the lowest energy state in a part of the parameter space which is consistent with the wave vectors found in La-214 compounds and BSCCO compounds but different from the wave vector found in YBa$_2$Cu$_3$O$_{6+x}$  at the same doping. However, what is more significant is the finding that the simple model Hamiltonian of Equation~\eref{eq:Heff2} can select lowest energy states with varying wave lengths and properties depending on the microscopic parameters. Specifically, the spin 
correlations in YBa$_2$Cu$_3$O$_{6+x}$ might be more strongly dictated by the Fermi surface of the doped holes, leading to different charge ordering wavevector,
which may allow one to reconcile the apparently quite different observed broken symmetries in this family of materials. In conclusion, our results support the view that 
despite system-specific differences, the  various symmetry-breaking phenomena can be driven by same driving force of local magnetic correlations.

\vspace{5mm}
\noindent{\bf Acknowledgements:} We thank Erez Berg, W. Buyers, J.C. Seamus Davis, Eduardo Fradkin, Marc-Henri Julien, Steve Kivelson, Kai Sun, John Tranquada  for helpful discussions. M.H.F was supported in part by Cornell Center for Materials Research with funding from the NSF MRSEC program (DMR-1120296) and by the Swiss Society of Friends of the Weizmann Institute of Science. E.-A.K was supported by the U.S. Department of Energy, Office of Basic Energy Sciences, Division of Materials Science and Engineering under Award DE-SC0010313. 
S.W. and A.P. acknowledge funding from NSERC of Canada.
\vspace{5mm}

\bibliographystyle{unsrt}
\bibliography{ref}

\appendix
\section{Diagonalization of the hole Hamiltonian}
\label{sec:app}
\subsection{Antiferromagnetic configuration}
For the long-range N\'eel antiferromagnetic order, {\it i.e.}, $S_i^z=\frac{1}{2}(-1)^{i_x+i_y}$, the Hamiltonian equation~\eref{eq:Heff2} becomes
\begin{eqnarray}\nonumber
\!\!\!\!\!\!\!\!\!\!\!\!\!\!\!\!\!\!\!\!\!\!\!\!\!\!\! \HH&=& \frac{t_a-t_b}{2} \sum_{\langle IiI'\rangle,s}\hat p_{Is}^{\dag}\hat p^{\phantom{\dag}}_{I' s}
+ \frac{t_a+t_b}{2}\sum_{\langle IiI'\rangle,s} s (-1)^{i_x + i_y}(\hat p_{I s}^{\dag}\hat p^{\phantom{\dag}}_{I' s}) 
- t_{pp} \sum_{\langle I I'\rangle,s}\hat p_{Is}^{\dag}\hat p^{\phantom{\dag}}_{I' s} \\
\!\!\!\!\!\!\!\!\!\!\!\!\!\!\!\!\!\!\!\!\!\!\!\!\!\!\!
&& -\frac{NJ_1}{2},
  \label{eq:HeffAFM}
\end{eqnarray}
where the exchange term between O and Cu atoms is zero due to the zero total spin component in the 
$z$ direction.
Opposite spin components are decoupled, therefore the hole part of the Hamiltonian can be written as
$\HH = \sum_{\bk,s}'\psi^{\dag}_{\bk s} \HH_{\bk s}\psi^{\phantom{\dag}}_{\bk s}$
with $\psi^{\dag}_{\bk s} = (p_{x \bk s}, p_{y \bk s}, p_{x \bk + \bQ s}, p_{y \bk + \bQ s})^{\dag}$, $\bQ=(\pi, \pi)$. 
Note that the sum $\sum_{\bk, s}'$ here only runs over the folded Brillouin zone.
The momentum part is given by 
\begin{equation}
  \HH_{\bk s}=\left(
    \begin{array}{cc}
      \HH_1(\bk) & s \HH_2(\bk) \\
      s\HH_2^{{\rm \dag}}(\bk) & \HH_1(\bk+\bQ) \\
    \end{array}
  \right),
  \label{eq:Hafm}
\end{equation}
where 
\begin{eqnarray}
\mathcal{H}_{1}(\bk)\!&\!=\!&\! 
  \left(
    \begin{array}{cc}
      (t_a \! -\! t_b) \cos k_x \!\!&\!\! 2 (t_a\!-\! t_b \!-\! 2 t_{pp}) \cos \frac{k_x}{2} \cos \frac{k_y}{2} \\
      2 (t_a\!-\! t_b \!-\! 2 t_{pp}) \cos \frac{k_x}{2} \cos \frac{k_y}{2} \!\!&\!\! (t_a-t_b) \cos k_y
    \end{array}
  \right) \!,\label{eq:h1}\\
  \HH_2(\bk)\!&\!=\!&\! (t_a+t_b)\!\left(
    \begin{array}{cc}
      i \sin k_x & 2 i \cos\frac{k_x}{2}\sin\frac{k_y}{2}\\
      2 i \sin \frac{k_x}{2}\cos \frac{k_y}{2} &  i \sin k_y
    \end{array}
  \right)\!\!.
  \label{eq:h2}
\end{eqnarray}
Diagonalization of the matrices leads to the band structure for spin up and down holes, which are degenerate. 

\subsection{ Spiral configurations}
The spiral spin pattern is given by $\bS_i=S(\cos \bQ\cdot{\bf r}_i ,\sin\bQ\cdot{\bf r}_i,0)$ and the Hamiltonian equation~\eref{eq:Heff2} can again be written in closed form in momentum space as
$\HH=\sum_{\bk}\psi_{\bk}^{\dagger}\mathcal{H}(\bk)\psi_{\bk}$, now with
$\psi_{\bk}^{\dagger}=(p_{x,\bk,-\bQ \ua},p_{y,\bk-\bQ\ua},p_{x,\bk\da},p_{y,\bk\da})^{\dagger}$, where
\begin{equation}
  \HH(\bk)=\left(
    \begin{array}{cc}
      \mathcal{H}_{1}(\bk-\bQ) & \mathcal{H}_{2}(\bk,\bQ) \\
      \mathcal{H}_{2}^{{\rm T}} (\bk,\bQ) & 
      \mathcal{H}_{1}(\bk) \\
    \end{array}
  \right)
\end{equation}
with $\HH_1(\bk)$ given in equation~\eref{eq:h1} and
\begin{equation}
\mathcal{H}_{2}(\bk,\bQ) \!=\! 2 S (t_a + t_b)
  \left(
    \begin{array}{cc}\!\!
      \cos (k_{x} \!-\! \frac{Q_x}{2}) \!+\! J_2 \cos \frac{Q_{x}}{2} \!\! &\!\! 2 \cos\frac{k_y}{2} \cos(\frac{k_x}{2} \!-\! \frac{Q_x}{2})\\
     2 \cos\frac{k_x}{2} \cos(\frac{k_y}{2} \!-\! \frac{Q_y}{2}) \!\! &\!\!  \cos (k_{y} - \frac{Q_y}{2}) \!+\! J_2 \cos \frac{Q_{y}}{2}\!\!\!
    \end{array}
  \right)
 \end{equation}
Depending on the direction of spiral propagation we consider a diagonal spiral 
with $\bQ=(\pi-\delta, \pi-\delta)$, a parallel spiral with respect to FM arrangement with $\bQ=(\pi-\delta, 0)$, and a parallel spiral with respect to AFM 
arrangement with $\bQ=(\pi-\delta, \pi)$. Figure~\ref{fig:patterns} summarizes the spiral patterns we considered.

\end{document}